\documentclass[5p,times]{elsarticle}
\usepackage{lineno,hyperref,multirow,graphicx}
\usepackage{multirow,graphicx}
\modulolinenumbers[5]

\journal{Elsevier}
\begin{document}
\begin{frontmatter}
\title{Evaluating user reputation in online rating systems via an iterative group-based ranking method}

\author[inst1]{Jian Gao\corref{cor1}}
\cortext[cor1]{Email addresses: gaojian08@hotmail.com (Jian Gao)}
\author[inst1,inst2]{Tao Zhou\corref{cor2}}
\cortext[cor2]{Email addresses: zhutou@ustc.edu (Tao Zhou)}

\address[inst1]{CompleX Lab, Web Sciences Center, University of Electronic Science and Technology of China, Chengdu 611731, People's Republic of China}
\address[inst2]{Big Data Research Center, University of Electronic Science and Technology of China, Chengdu 611731, People's Republic of China}

\begin{abstract}
Reputation is a valuable asset in online social lives and it has drawn increased attention. How to evaluate user reputation in online rating systems is especially significant due to the existence of spamming attacks. To address this issue, so far, a variety of methods have been proposed, including network-based methods, quality-based methods and group-based ranking method. In this paper, we propose an iterative group-based ranking (IGR) method by introducing an iterative reputation-allocation process into the original group-based ranking (GR) method. More specifically, users with higher reputation have higher weights in dominating the corresponding group sizes. The reputation of users and the corresponding group sizes are iteratively updated until they become stable. Results on two real data sets suggest that the proposed IGR method has better performance and its robustness is considerably improved comparing with the original GR method. Our work highlights the positive role of users' grouping behavior towards a better reputation evaluation.
\end{abstract}

\begin{keyword}
Rating systems\sep Reputation evaluation\sep Ranking method \sep Spamming attack \sep Iterative refinement
\end{keyword}

\end{frontmatter}
\linenumbers
\nolinenumbers

\section{Introduction}
At the age of Internet, individual reputation plays the role of fundamental blocks in building up online ecosystems, especially in the filed of e-commerce \cite{Resnick2000rep,Standifird2001}. Meanwhile, new challenges arise that how to create and maintain reputation in online communities? To better uncover objects' true quality, many platforms implement online rating systems, e.g. Amazon, eBay, Taobao, MovieLens, where users can give their feedbacks by assigning ratings to objects \cite{Linyuan2012,Bobadilla2013}. The ratings provide a direct measure of reputation for the objects and further affect users' decisions \cite{Josang2007,Bente2012to,Zhao2013,Yu2015}. Usually, high ratings result in high sales whereas low ratings play the opposite role. As a result, to extract credible information from these abundant feedbacks is becoming a major challenge since noisy ratings are widely existed in practical systems \cite{Muchnik2013,Yang2012,Toledo2015}. For example, some users may give unreasonable ratings due to their poor judgement \cite{Chirita2005,Xie2012}, and some others may purposefully guide public choices by giving maximal/minimal ratings \cite{Zeng2012re,Benevenuto2009}. These noisy ratings can harm the effectiveness of online rating systems and affect the accuracy of the obtained information \cite{Mukherjee2011,Lin2014to,Sun2012}. Therefore, how to measure users' credibility, filter out untrusted users and ensure reliability of online rating systems are becoming urgent tasks \cite{liu2014new,Zhang2012cj,Lim2010de}.

To cope with these concerns, online reputation systems are introduced \cite{Ling2013a,Hung2012}. These systems are capable of decision support for Internet-mediated services and help to maintain the healthy development of online rating systems and recommender systems. As the core of reputation systems, a variety of user reputation evaluation methods have been proposed \cite{Li2012ro,Khosravifar2012}, where each user is assigned with a reputation value based on their rating behaviors \cite{Fujimura2003,Liu2015rank}. Typically, these previous methods can be divided into three categories:
\begin{itemize}
  \item Network-based methods. As online rating systems can be described by bipartite networks \cite{Shang2010}, the reputation for users can be calculated by many existing networked ranking methods such as PageRank \cite{Yamamoto2004}, LeaderRank \cite{lv2011le}, mass diffuse \cite{Zhang2007re,Zhou2007bi} and heat conduction \cite{Zhang2007he}. In these methods, a user's reputation is measured by the amount of resources that the user receives in the resource-allocation processes. Although these methods are very efficient, they suffer from rating noises and thus have limited performance \cite{Li2012ro}. As a result, these methods are not suitable for user reputation evaluation in bipartite networks.
  \item Quality-based methods. Underlying an assumption that each object has a most objective rating that best reflects its quality \cite{Tian2012}, the quality-based methods measure a user's reputation by the difference between the rating values and the estimated objects' quality values \cite{Liao2014towards}. These methods include iterative refinement (IR) method \cite{Laureti2006}, an improved IR method \cite{De2007iterative}, correlation-based ranking (CR) method \cite{Zhou2011a}, reputation redistribution ranking (RR) method \cite{Liao2014} and the other seven methods \cite{Li2012ro,Liu2015rank}. These aforementioned methods are well-performed in user reputation evaluation, however, some of them may not converge and some others are not robust to spamming attacks \cite{Li2012ro,Allahbakhsh2015}. More importantly, due to the fact that the online rating system is fundamentally a socialized information collection platform, one object should accept multiple reasonable ratings \cite{Tian2012} since the ratings are subjective and can be affected by users' background and some other factors \cite{Muchnik2013,Yang2012}. Therefore, the underlying assumption of quality-based methods is worthy of scrutiny.
  \item  Group-based method. Recently, a group-based ranking (GR) method is proposed, in which users are grouped based on their rating similarities, and users' reputation is calculated by the corresponding group sizes \cite{Gao2015}. Users are assigned with high reputation if they always fall into large rating groups. This method is free from the assumption of the quality-based methods and it has better performance in evaluating user reputation on data sets with spamming attacks. However, the method is not robust for plenty of large-degree spammers as it's one-step process and the ratings are evenly contributed in calculating the corresponding group sizes regardless of users' reputation.
\end{itemize}

In this paper, we propose an iterative group-based ranking (IGR) method by introducing an iterative reputation-allocation process into the original GR method. Specifically, ratings from users with high reputation are assigned with higher weights in calculating the corresponding group sizes. Both the user reputation and the group sizes are iteratively updated until they become stable. This method is partially inspired by the GR method \cite{Gao2015}, the original resource-allocation process \cite{Zhou2007bi,Ou2007}, and the HITS algorithm with iterative refinement procedure \cite{Kleinberg1999}. When tested on two real data sets (MoiveLens and Netflix) with artificial spammers, the proposed IGR method has excellent performance in evaluating user reputation and its robustness in resisting a large number of spammng attacks is considerably improved compared with the original GR method. Further, provided some insights on the mechanism and analyzed the characteristics of these methods. Results suggest that IR method remarkably prefers large-degree users, CR and RR methods have no obvious degree preference, and GR and IGR methods slightly prefer small-degree users. Our work provides a further understanding on some reputation evaluation methods and highlights the significance of considering users' grouping behaviors in designing better reputation systems.

\begin{figure*}[!ht]
\centering
\includegraphics[width=185mm]{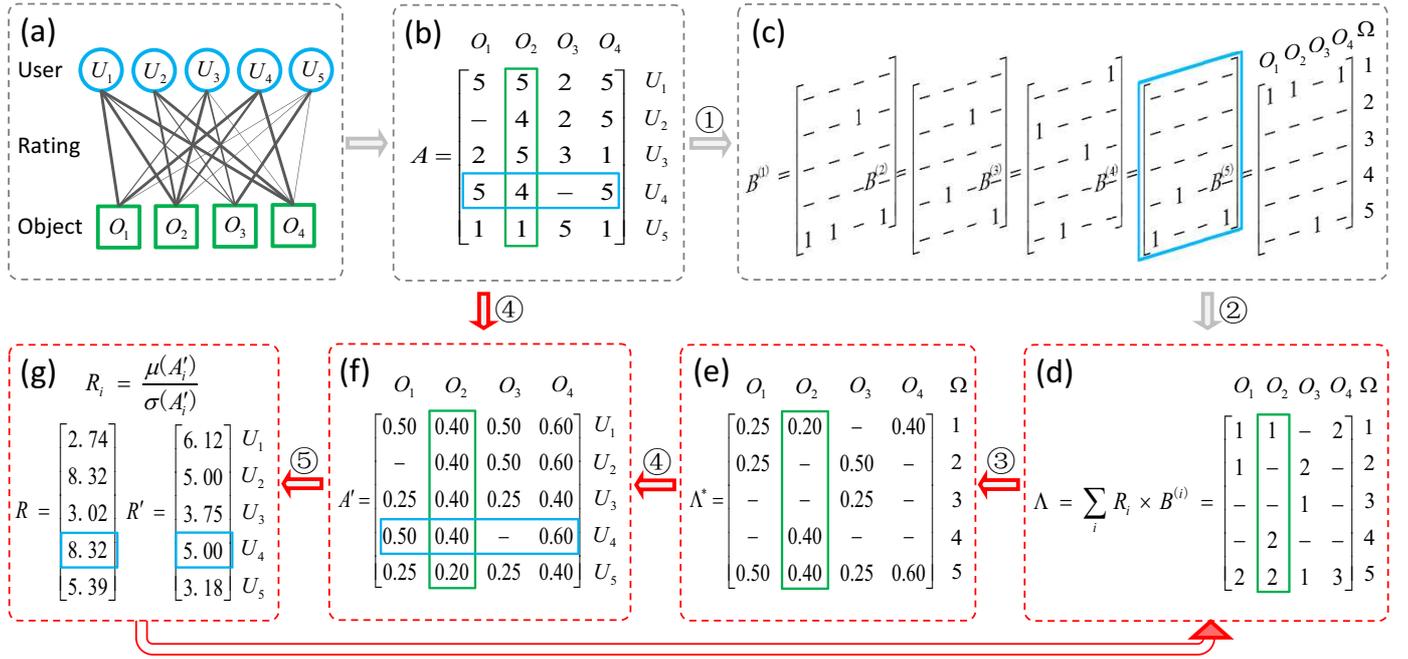}
    \caption{Illustration of the IGR method. The number besides the arrow marks the order of the procedure. The symbol ``-" in matrixes stands for a non-value, which should be ignored in the calculation. (a) The original weighed bipartite network, $G$. (b) The corresponding rating matrix, $A$. The row and column correspond to users and objects, respectively. (c) The rating-object matrix for user $i$, $B^{(i)}$. Taking $U_{4}$ as an example (blue horizontal box in (b)), $B^{(4)}_{5,1}=B^{(4)}_{4,2}=B^{(4)}_{5,4}=1$. (d) The reputation-weighted group size matrix, $\Lambda$. Taking $O_{2}$ as an example (green vertical box in (b)), $\Lambda_{4,2} = R_{2}\times B^{(2)}_{4,2}+R_{4}\times B^{(4)}_{4,2}$=2. (e) The rating-rewarding matrix, $\Lambda^{\ast}$, constructed by normalizing $\Lambda$ by column, e.g, $\Lambda^{\ast}_{4,2}=2/(1+2+2)=0.40$. (f) The rewarding matrix, $A'$, obtained by mapping matrix $A$ referring to $\Lambda^{\ast}$, e.g. $A'_{4,2}=0.40$. (g) The reputation of users, $R$. $R'$ is temporal reputation in the previous iteration step. In IGR method, $\Lambda$ and $R'$ are iteratively updated according to (d), (e), (f) and (g), as indicated by the red arrows. Finally, a stable reputation $R$ is obtained.}
    \label{fig:IGR}
\end{figure*}

\section{Methods}
We first introduce some basic notations for the user reputation evaluation methods. The online rating system can be naturally described by a weighed bipartite network $G=\{U, O, E\}$, where $U=\{U_{1}, U_{2}, ..., U_{m}\}$, $O=\{O_{1}, O_{2}, ..., O_{n}\}$ and $E=\{E_{1}, E_{2}, ..., E_{l}\}$ are sets of users, objects and ratings (see Fig.~\ref{fig:IGR}a for an illustration), respectively. Here, we use Greek and Latin letters, respectively, for object-related and user-related indices to distinguish them. The degree of a user $i$ and an object $\alpha$ are denoted as $k_i$ and $k_{\alpha}$, respectively. Considering a discrete rating system, the bipartite network can be represented by a rating matrix $A$, where the element $A_{i\alpha} \in \Omega=\{\omega_{1}, \omega_{2}, ..., \omega_{z}\}$ is the weight of the link connecting user $i$ and object $\alpha$, with $A_{i\alpha}$ being equal to the corresponding rating value (see Fig.~\ref{fig:IGR}b). In a reputation system, each user $i$ will be assigned with a reputation value, which is denoted as $R_{i}$. In the following, we will briefly introduce the proposed user reputation evaluation method.

\subsection{Group-based ranking methods}
The iterative group-based ranking (IGR) method and the original group-based ranking (GR) method are based on the same framework. Thus, we mainly introduce the IGR method. After the initial configuration that each user $i$ has equal reputation, e.g., $R_{i} = 1$, the IGR method works as follows.

Firstly, for user $i$, the rating vector $A_{i}$ is mapped to a rating-object matrix $B^{(i)}$, whose element $B^{(i)}_{s\alpha}$ is defined as
\begin{equation}
\label{eq:B}
B^{(i)}_{s\alpha}=\left\{
\begin{array}{lcl}
 1 & & \mbox{if $A_{i\alpha}=\omega_{s}$}\\
 - & & \mbox{otherwise}
\end{array}
,
\right.
\end{equation}
where the symbol ``-" stands for a non-value, which should be ignored in the calculation (the same below). In this way, users are grouped by their ratings, namely, users who give the same rating $\omega_{s}$ to object $\alpha$ belong to the group $\Gamma_{s\alpha}$. Mathematically, the group is defined as $\Gamma_{s\alpha}=\{U_{i}|B^{(i)}_{s\alpha}=1\}$. Obviously, user $i$ belongs to $k_{i}$ different groups.

Secondly, based on the intuition that a user with poor reputation should have less chance in forming big groups, we calculate the size of group $\Gamma_{s\alpha}$ by considering both the rating-object matrix $B^{(i)}$ and users' reputation $R_{i}$. Mathematically, the weighted group size $\Lambda_{s\alpha}$ is defined as
\begin{equation}
\label{eq:Lam}
\Lambda_{s\alpha}=\sum^{m}_{i=1}R_{i}\cdot B^{(i)}_{s\alpha},
\end{equation}
where $m$ is the number of users. Then, a rating-rewarding matrix $\Lambda^{\ast}$ is established by normalizing matrix $\Lambda$ by column. Mathematically, $\Lambda^{\ast}_{s\alpha}=\Lambda_{s\alpha}/k_{\alpha}$.

Thirdly, referring to the rating-rewarding matrix $\Lambda^{\ast}$, the original rating matrix $A$ is mapped to a rewarding matrix $A'$. Specifically, the rewarding $A'_{i\alpha}$ that user $i$ obtains from the rating $A_{i\alpha}$ is defined as
\begin{equation}
\label{eq:A2}
A'_{i\alpha}=\left\{
\begin{array}{lcl}
 \Lambda^{\ast}_{s\alpha} & & \mbox{if $A_{i\alpha}=\omega_{s}$}\\
 - & & \mbox{otherwise}
\end{array}
.
\right.
\end{equation}

Finally, the reputation is re-allocated to all users according to their rewarding vectors. On the one side, if the average of a user's rewarding is small, most of his ratings must be deviated from the majority, indicating his/her poor reputation. On the other side, if the rewarding varies largely, he/she is also untrustworthy for the unstable rating behavior. Based on these intuitions, the reputation $R_i$ for user $i$ is calculated as
\begin{equation}
\label{eq:R}
    R_{i}=\frac{\mu(A'_{i})}{\sigma(A'_{i})}=\frac{(\sum_{\alpha\in O_i}A'_{i\alpha})^2}{\sum_{\alpha\in{O_i}}(k^{2}_{i}A'_{i\alpha}-k_{i}\sum_{\alpha\in{O_i}}A'_{i\alpha})^2},
\end{equation}
where $\mu$ and $\sigma$ are mean value and standard deviation, respectively.

In IGR, the reputation $R$ and the group size $\Lambda$ are iteratively updated according to Eqs.~(\ref{eq:Lam}), (\ref{eq:A2}) and (\ref{eq:R}) until the change of the reputation $|R-R'|=\sum_{i}(R_{i}-R'_{i})^2/m$ is smaller than the threshold value $\Delta=10^{-4}$. Here, $R'$ denotes the reputation vector at the previous iteration step. Note that, when there is no iteration, IGR degenerates to the original GR. A visual representation of the IGR method is shown in Fig.~\ref{fig:IGR}.

\subsection{Quality-based ranking methods}
Quality-based ranking methods have an underlying assumption that each object $\alpha$ is associated with a most objective rating that best reflects its true quality $Q_{\alpha}$. As it's really hard to tell the true quality of objects, as an alternative, the estimated quality $\hat{Q}_{\alpha}$ of object $\alpha$ is usually used, which is defined as the objects' weighted average rating. Mathematically, it reads
\begin{equation}
\label{eq:QIR}
\hat{Q}_{\alpha}= \frac{\sum_{i\in{U_{\alpha}}}{R_{i}A_{i\alpha}}}{\sum_{i\in{U_{\alpha}}}{R_i}},
\end{equation}
where $U_{\alpha}$ is the set of users who have rated object $\alpha$, and $A_{i\alpha}$ is the rating to object $\alpha$ from user $i$ with reputation $R_{i}$. Here, we consider three representative quality-based ranking methods, namely, iterative refinement (IR) \cite{Laureti2006}, correlation-based ranking (CR) \cite{Zhou2011a}, reputation redistribution ranking (RR) \cite{Liao2014}.

The IR method calculates the user reputation and object quality in an iterative way. Specifically, a user's reputation is inversely proportional to the difference between the rating vector and the corresponding objects' estimated quality vector. Mathematically, the difference is defined as
\begin{equation}
\label{eq:dIR}
  f_i = \frac{1}{k_{i}} \sum_{\alpha\in {O_{i}}} (A_{i\alpha} - \hat{Q}_{\alpha})^2,
\end{equation}
where $\hat{Q}_{\alpha}$ is the estimated quality value of object $\alpha$. Initially, all users have the same reputation, e.g., $R_{i}=1$. Then, the reputation of user $i$ is iteratively updated according to
\begin{equation}
\label{eq:RIR}
  R_i = \left(f_i + \varepsilon \right)^{-\beta},
\end{equation}
where $\beta$ is a tunable parameter, whose optimal value is around $\beta=1$ \cite{Liao2014}. The iteration goes according to Eqs.~(\ref{eq:QIR}), (\ref{eq:dIR}) and (\ref{eq:RIR}) until both $\hat{Q}_{\alpha}$ and $R_i$ converge.

As CR and RR methods are based on the same framework, in the following, only RR is introduced. In RR, each user $i$ is initially with reputation $R_{i} = k_{i}/n$, which can be essentially seen as the user's activity. The estimated quality of objects is calculated by Eq.~(\ref{eq:QIR}). To obtain the reputation $R_i$ for user $i$ in a step, a so-called temporal reputation $TR_i$ is calculated, which is the Pearson correlation coefficient between the rating vector $A_{i}$ and the estimated objects' quality vector $Q_{i}$. Mathematically, $TR_i$ is defined as
\begin{equation}
TR_i=\frac{1}{k_{i}}\sum_{\alpha\in{O_i}}{\left(\frac{A_{i\alpha}-\mu(A_i)}{\sigma(A_i)}\right)}{\left(\frac{\hat{Q}_{\alpha}-\mu(\hat{Q}_i)}{\sigma(\hat{Q}_i)}\right)},
\end{equation}
where $\mu$ and $\sigma$ are functions of mean value and standard deviation, respectively. If $TR_i$ is smaller than 0, $TR_i$ is reset as 0, leading $TR_i$ being in the range $[0,1]$. Then, the reputation $R_{i}$ is obtained by nonlinearly redistributing $TR_i$ via
\begin{equation}
R_i=TR_i^{\theta}\frac{\sum_j TR_j}{\sum_j TR_j^{\theta}},
\end{equation}
where $\theta$ is a tunable parameter. Note that RR degenerates to CR when $\theta=1$. In each step, both $\hat{Q}_{\alpha}$ and $R_{i}$ are updated until the change of the estimated quality $|{\hat{Q}-\hat{Q}'}|= \sum_{\alpha\in{O}}{(\hat{Q}_{\alpha}-{\hat{Q}'_{\alpha}})^2}/n$ is smaller than a threshold value $\Delta=10^{-4}$. Here, $\hat{Q}'$ denotes the vector of objects' qualities in the previous step, and the parameter $\theta$ is set as its optimal value $\theta=3$ \cite{Liao2014}.

\section{Data and metric}
\subsection{Real rating data}
We consider two commonly used data sets in online rating systems, namely, MovieLens and Netflix. Both of the two data sets contain ratings on movies based on a 5-point rating scale with 1 being the worst and 5 being the best. MovieLens data set is provided by GroupLens project at University of Minnesota (www.grouplens.org). Herein, we only use a small subset, which is sampled and extracted from the original data with the constraint that each user has at least 20 ratings and the movies are rated by at least one of these users. In the subset, 100000 ratings are given by 943 users to 1682 movies. Netflix is a huge data set released by the DVD rental company Netflix for its Netflix Prize contest (www.netflixprize.com). We extracted a small data set by random choosing 3000 users who have at least 20 ratings and took all 2779 movies that rated by at least one of these users. Finally, there are 197248 ratings in the Netflix data set. Compared with Netflix, MovieLens has larger average user degree, smaller average object degree and higher sparsity. The basic statistics of data sets are summarized in Table~\ref{tab:stat}.

\begin{table}
\caption{Some basic characteristics of real data sets. $m$ is the number of users, $n$ is the number of objects, $\langle k_U\rangle$ is the average degree of users, $\langle k_O\rangle$ is the average degree of objects, and $S=l/mn$ is the sparsity of the bipartite network, where $l$ is the number of all ratings.}
\label{tab:stat}
\begin{tabular*}{0.5\textwidth}{@{\extracolsep{\fill}}llllll}
    \hline
    Data set & $m$ & $n$ & $\langle k_U\rangle$ & $\langle k_O\rangle$ & $S$ \\
    \hline
    MovieLens & 943 & 1682 & 106 & 60 & 0.0630\\
    Netflix & 3000 & 2779 & 66 & 71 & 0.0237\\
    \hline
\end{tabular*}
\end{table}

\subsection{Artificial rating data}
To test the performance of different ranking methods, one way is to calculate the ranks of all users and compare them with the ground truth. However, in practice, we are unable to know the ground true ranks of users in advance. As an alternative, we manipulate the real data set by adding artificial spammers and test to what extent these spammers can be detected by a ranking method. In fact, two types of distorted ratings, namely, malicious ratings and random ratings are widely found in real online rating systems \cite{Ricci2011,Jindal2008}. The malicious ratings are from spammers who always gives minimum (maximum) allowable ratings to push down (up) certain target objects. The random ratings mainly come from test engineers or some naughty users who give meaningless ratings randomly.

As real spammers are unknown, to generate artificial rating data sets, we add either type of artificial spammers (i.e. malicious or random) at one time into the original data. In the implementation, we randomly select $d$ users and turn them into spammers by replacing their original ratings with distorted ratings: (i) integer 1 or 5 with the same probability (i.e., 0.5) for malicious spammers, and (ii) random integers in $\{ 1,2,3,4,5 \}$ for random spammers. Thus, the ratio of artificial spammers is $p=d/m$, where $m$ is the number of all users.

\subsection{Evaluation metric}
We apply two widely used metrics to evaluate the performance of ranking, namely, recall \cite{Herlocker2004} and AUC (the area under the ROC curve) \cite{Hanley1982}. The recall only focuses on the top-$L$ ranks and its value measures to what extent the spammers can be ranked at the top. Mathematically, the recall is defined as
\begin{equation}
\label{eq:RC}
    R_{c}(L)=\frac{d'(L)}{d},
\end{equation}
where $d'(L)\leq d$ is the number of detected artificial spammers in the top-$L$ ranking list. In the following experiments, the length of ranking list is set as $L=d$, at which setting recall is equivalent to another accuracy metric named precision \cite{Herlocker2004}. Larger value of $R_{c}$ indicates higher accuracy of the ranking.

Next, we introduce the $L$-independent metric AUC. Given the ranks of all users, the value of AUC value can be essentially seen as the probability that the reputation of a randomly chosen spammer is lower than that of a randomly chosen normal user (non-spammer) \cite{Linyuan2012}. To calculate AUC, at each time a pair of spammer and normal user are picked and their reputations are compared. If among $N$ independent comparisons, there are $N'$ times the spammer has a lower reputation and $N''$ times they have the same reputation, the AUC value is defined as
\begin{equation}
\label{eq:AUC}
    AUC=\frac{N' + 0.5N''}{N}.
\end{equation}
The value of AUC should be about 0.5 if all users and spammmers are ranked randomly. Therefore, the more the value of AUC exceeds 0.5, the better the ranking method performs.

\subsection{Self-consistency metric}
For the reputation evaluation methods, there is an intuition that a user of higher rating error should have a lower reputation or vice versa. That is to say, for a well-performed method, the reputation should be negatively correlated with the rating error. Here, the rating error of a user refers to the degree of deviation after comparing the rating $A_{i}$ and the estimated objects' quality $\delta_{i}$. Mathematically, for user $i$, the rating error $\delta_{i}$ is defined as
\begin{equation}
\label{eq:DE}
    \delta_{i}=\frac{\sum_{\alpha\in{O_i}}|A_{i\alpha}-\hat{Q}_{\alpha}|}{k_{i}},
\end{equation}
where $O_i$ is the set of objects being rated by user $i$, and $\hat{Q}_{\alpha}=\sum_{i \in U_{\alpha}}A_{i\alpha}/k_{\alpha}$ is the average rating that object $\alpha$ receives. In fact, the correlation between $\delta_{i}$ and $R_i$ measure the self-consistent of a ranking method as $\delta_{i}$ depends on $\hat{Q}$ and $\hat{Q}$ depends on $R_i$ alternately. The higher the correlation is, the more self-consistent the method is.

\begin{figure*}
\centering
\includegraphics[width=185mm]{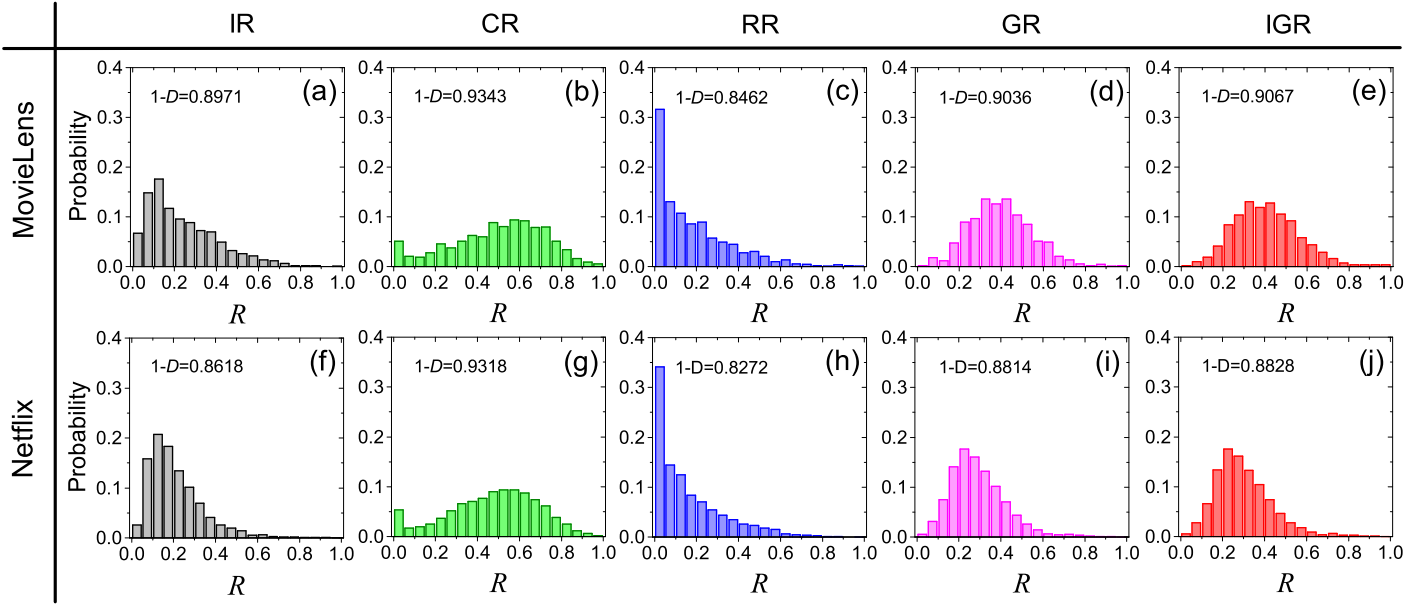}
    \caption{The probability distribution of users' reputation after applying different reputation evaluation methods on the two real online rating data sets, MovieLens and Netflix. Subfigures (a), (b), (c), (d) and (e) are for MovieLens; subfigures (f), (g), (h), (i) and (j) are for Netflix. $R$ is the reputation of users. $1-D$ is the Simpson's index of diversity.}
    \label{fig:Rep}
\end{figure*}

\section{Results}
\subsection{Reputation evaluation}
First, we consider the probability distribution of users' reputation after applying the reputation evaluation methods on the real online rating data sets. Results are shown in Fig.~\ref{fig:Rep}. It can be seen that in IR the reputation is Possion-like distributed whereas in CR, GR and IGR the reputation is normal-like distributed. By contrast, in RR the reputation is exponential-like distributed, which is remarkably different as the reputation of most users is zero (see Figs.~\ref{fig:Rep}c and \ref{fig:Rep}h). To quantify the diversity of all users' reputation from the probability distribution, we calculate the Simpson's index of diversity, which is denoted as $1-D$ \cite{Simpson1949}. Higher value of $1-D$ suggest more distinguishable of the obtained reputation. In CR, the values of $1-D$ are highest as 0.9343 and 0.9318 for MovieLens and Netflix, respectively. In GR and IGR, the values of $1-D$ are nearly the same, which are around 0.90 and 0.88 for MovieLens and Netflix, respectively. In RR, the values of $1-D$ are the lowest, suggesting that the reputation of users' in RR is the least distinguishable. Actually, the reputation a well-performed reputation evaluation method assigns should be distinguishable, and CR, GR and IGR perform better.

\begin{figure*}
\centering
\includegraphics[width=185mm]{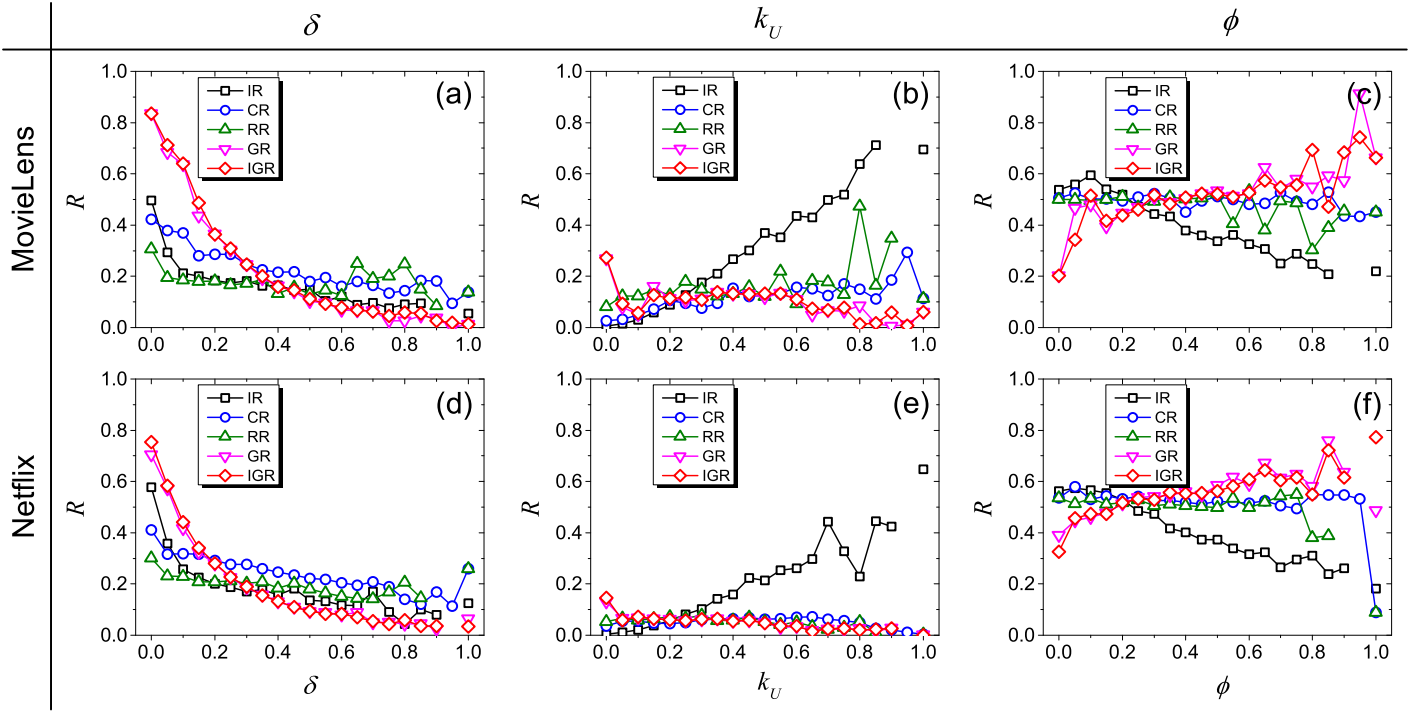}
    \caption{The relation between $R$ and $\delta$, $k_{U}$ and $\phi$, respectively. Subfigures (a), (b) and (c) are for MovieLens; subfigures (d), (e) and (f) are for Netflix. $\delta$ is the rating error of users, $k_{U}$ is the degree of users, and $\phi$ is the degree of trend following. For comparison, $\delta$, $k_{U}$ and $\phi$ are respectively normalized. As the three normalized indicators are continuous, we respectively divide them into bins with the length 0.05 and then evaluate the mean reputation of users in the same bins.}
    \label{fig:Corr}
\end{figure*}

Then, in Figs.~\ref{fig:Corr}a and \ref{fig:Corr}d, we show the relation between $\delta$ and $R$, i.e. the self-consistency, for different methods. We note that GR and IGR both assign a high reputation to users of low rating errors and a stably low reputation to users of high rating errors. By contrast, the other three quality-based ranking methods, i.e., IR, CR and RR, are not stable in dealing with users of high rating errors, as indicated by high variation of $R$ when $\delta$ is large. To quantify the relation, we additionally calculate the Pearson correlation coefficient $\rho$ between $R$ and $\delta$. Results are shown in the first row of Table~\ref{tab:Corr}. The values of $\rho$ are respectively $-0.8166$ and $-0.8201$ ($-0.7353$ and $-0.7629$) for GR and IGR in MovieLens (Netflix) data set. The highest negative correlations suggest the best self-consistent of GR and IGR in user reputation evaluation.

\begin{table*}[t]
\centering
\caption{Pearson correlation coefficient $\rho$ between the reputation $R$ and the rating error $\delta$, the degree of users $k_{U}$ and the degree of trend following $\phi$, respectively. The highest correlation coefficients in each row are emphasized in bold.}
\label{tab:Corr}
\begin{tabular*}{185mm}{@{\extracolsep{\fill}}c|rrrrr|rrrrr}
\hline
\multicolumn{1}{c|}{\multirow{2}{*}{Metrics}} & \multicolumn{5}{c|}{MovieLens} & \multicolumn{5}{c}{Netflix}\\
\cline{2-11}
    & IR & CR & RR & GR & IGR & IR & CR & RR & GR & IGR\\
\hline
$\rho(\delta, R)$ & -0.4471 & -0.4537 & -0.3189 & -0.8166 & \textbf{-0.8201} & -0.4640 & -0.3926 & -0.2812 & -0.7353 & \textbf{-0.7629}\\
$\rho(k_{U}, R)$ & \textbf{0.8759} & 0.2318 & 0.1719 & -0.0519 & -0.0419 & \textbf{0.7868} & 0.0538 & 0.0040 & -0.0950 & -0.0904\\
$\rho(\phi, R)$ & \textbf{-0.4746} & -0.0244 & -0.0287 & 0.2141 & 0.2048 & \textbf{-0.3793} & -0.0428 & -0.0569 & 0.2368 & 0.2157\\ 		
\hline
\end{tabular*}
\end{table*}

We next consider the effect of user degree $k_{U}$ on determining the corresponding reputation $R$ under different ranking methods. Figs.~\ref{fig:Corr}b and \ref{fig:Corr}e show the relations between $k_{U}$ and $R$. It is worthy noticing that $R$ in IR is positively correlated with $k_{U}$ as the correlation is 0.8759 and 0.7868 for MovieLens and Netflix, respectively. In fact, the degree $k_{U}$ can be essentially seen as a user's activity. Thus, the result indicates that IR prefers users with high activity as it gives a higher reputation to active users than inactive ones. By contrast, for the other four methods, there is no obvious degree preference as the correlations are all around 0 (see the second row of Table~\ref{tab:Corr}). The main reason for these observations is that $R$ in IR is inversely proportional to the least mean square of the difference between $A_{i\alpha}$ and $\hat{Q}_{\alpha}$. As the difference is degree-dependent, in IR, large-degree users get a higher reputation in the iteration. While CR and RR calculate the correlation and GR and IGR calculate the mean and standard deviation, which are all independent of the user degree. In practice, there is another understanding of such positive correlation for IR. The user degree can be roughly seen as a reflection of buyers' experiences. Users of larger degree receive more information and they are experienced. Hence, it can be roughly considered that large degree users have better judgement and their reputation should be higher. However, the straightforward index is not enough to deal with the problem as it's hard to dig out large degree spammers.

\begin{figure*}[!t]
\centering
\includegraphics[width=140mm]{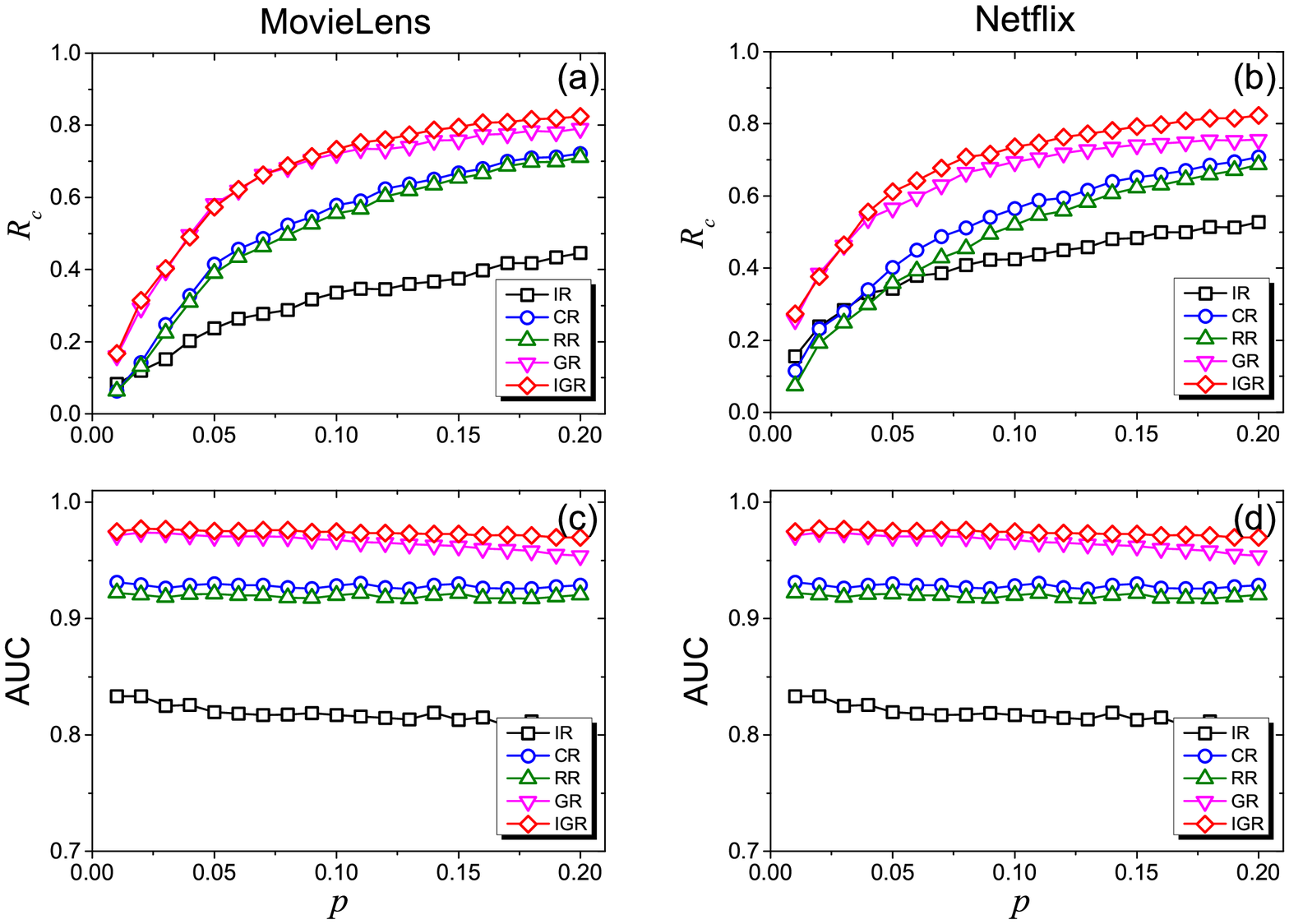}
    \caption{Performance of different methods on data sets with random spamming. Subfigures (a) and (b) are for $R_{c}$; subfigures (c) and (d) are for AUC. The parameter $p$ is the ratio of random spammers. Results are averaged over 100 independent realizations.}
    \label{fig:PerfRD}
\end{figure*}

Further, we study how the degree of trend following affects the reputation evaluation. The so-called degree of trend following measures to what extent a user would like to collect objects of high popularity. Usually, the popularity of an object is represented by its degree. Hence, a user's degree of trend following, denoted as $\phi$, can be calculated as the average degree of objects that rated by the user. Mathematically, it reads
\begin{equation}
\label{eq:KOU}
    \phi_{i}=\frac{\sum_{\alpha \in O_{i}}k_{\alpha}}{k_{i}},
\end{equation}
where $O_i$ is the set of objects that rated by user $i$, $k_i$ is the degree of user $i$, and $k_{\alpha}$ is the degree of object $\alpha$. The relations between the user reputation $R$ and the degree of trend following $\phi$ are shown in Figs.~\ref{fig:Corr}c and \ref{fig:Corr}f. It can be seen that $R$ in IR is negatively correlated with $\phi$ as the values of $\rho$ are $-0.4746$ and $-0.3793$ for MovieLens and Netflix, respectively (see the third row of Table~\ref{tab:Corr}). In GR and IGR, $R$ is weak positively correlated with $\phi$ as the value of $\rho$ is around 0.2. In CR and RR, the value of $\rho$ is around 0, indicating that $R$ is almost independent of $\phi$. To better understand these observations, we focus on the mechanisms of these methods. In IR, the ratings from a user of larger $\phi$ have less chance in dominating the corresponding object's quality, which finally results in the user's lower reputation. In GR and IGR, a lager $\phi$ ensures a stabler grouping, which results in a user's higher reputation. For a more intuitive understanding, we consider the real meaning of the differences among the correlation coefficients. Users who always buy things of high popularity have public taste and the information they receive is popular to audience. Thus, it's much harder for them to get higher reputation compared with the users who have their unique taste and richer information in IR. By contrast, users of larger degree with trend following have better grouping behavior in collecting objects and they should have higher reputation in GR and IGR.

\subsection{Random spamming analysis}
To evaluate the performance of different methods in resisting random spamming, we first generate artificial data sets with random spammers and then calculate $R_c$ and AUC accordingly. Results are shown in Fig.~\ref{fig:PerfRD}. When focusing on the top ranks, indicated by the value of $R_c$ in Figs.~\ref{fig:PerfRD}a and \ref{fig:PerfRD}b, GR and IGR both have the best performance, and IGR is more robust than GR. CR is on a par with RR, and they both outperform IR. Further, we note that the value of $R_{c}$ increases as $p$ increases. Specifically, the value of $R_{c}$ has a rapid growth when $p$ is approaching a value around 0.05. Afterwards, the value of $R_{c}$ becomes stable. The result suggests that there are some real random spammers in the original rating data sets, and the ratio is about 0.05. When focusing on the overall performance, indicated by the values of AUC in Figs.~\ref{fig:PerfRD}c and \ref{fig:PerfRD}d, GR and IGR remarkably outperform the other methods by giving a robust AUC value around 0.96. CR and RR are slightly inferior as the AUC value is about 0.92. For IR, the AUC value is significant lower, indicating its limited performance. In short, group-based methods outperform the quality-based methods in resisting random spamming.

\begin{figure*}[t]
\centering
\includegraphics[width=185mm]{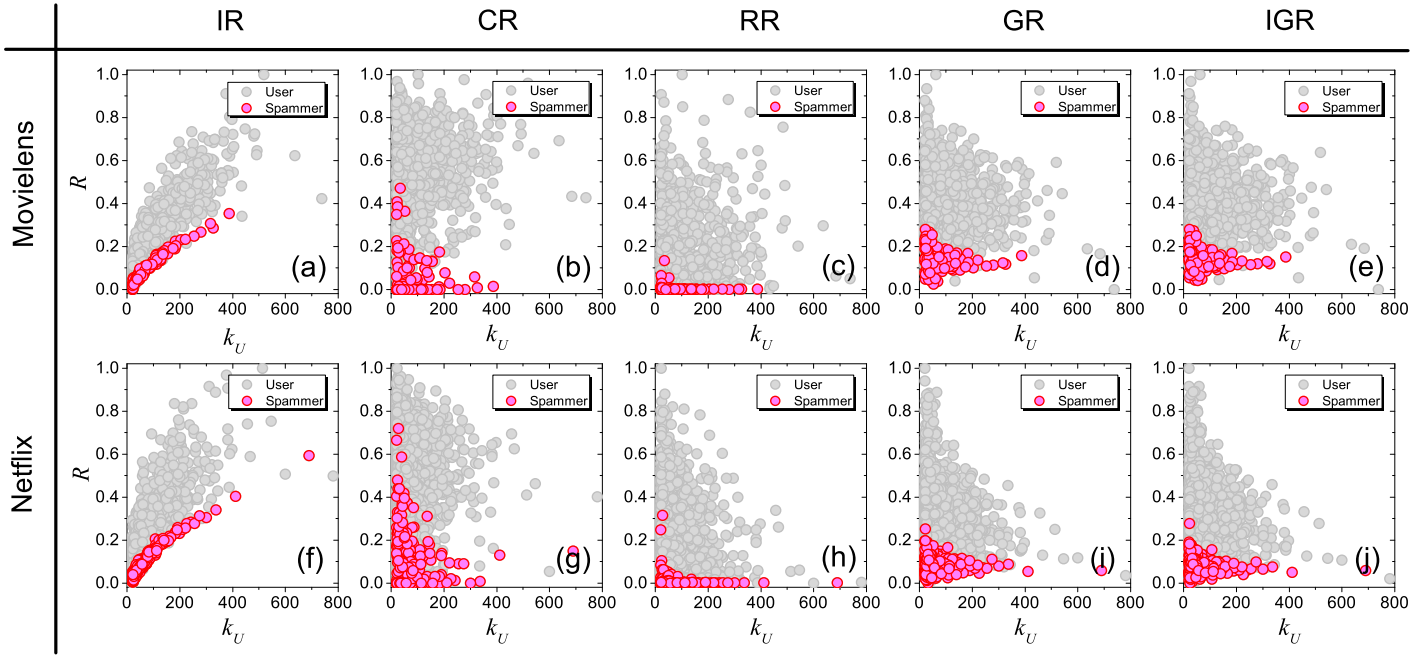}
    \caption{The relation between $R$ and $k_{U}$. $R$ is the reputation of users, obtained by applying different methods on data sets with random spamming. $k_{U}$ is the degree of users. The data points colored gray and pink stand for normal users and random spammers, respectfully. The parameter is set as $p=0.1$. Results in each subfigures are for one realization.}
    \label{fig:DegRD}
\end{figure*}

For a more intuitive understanding of how different methods work in resisting random spamming, in Fig.~\ref{fig:DegRD}, we show the effect of the user degree on reputation evaluation in parameter spaces ($R$, $k_{U}$). In can be seen that $R$ is positively correlated with $k_{U}$ in IR. Hence, for users with close degree, IR can accurately distinguish spammers from normal users as shown in Figs.~\ref{fig:DegRD}a and \ref{fig:DegRD}f. Despite of this, IR gives a lower reputation to many users (see Figs.~\ref{fig:Rep}a and \ref{fig:Rep}f) but a relatively higher reputation to spammers with large degree, which results in its poor performance. Meanwhile, CR gives all users (especially some small-degree spammsers) a relatively higher reputation, indicated by most of dots being in the middle and top of Figs.~\ref{fig:DegRD}c and \ref{fig:DegRD}h. In other words, the mean of all users' reputation in CR is relatively higher (see Figs.~\ref{fig:Rep}b and \ref{fig:Rep}g). By contrast, RR over limits all users reputation, as indicated by most dots being in the bottom of Figs.~\ref{fig:DegRD}c and \ref{fig:DegRD}h, although it gives most spammers a lower reputation. In RR, a lot of users have zero reputation (see Figs.~\ref{fig:Rep}c and \ref{fig:Rep}h), which results in a high false positive rate in spam detection. GR and IGR both slightly prefer small-degree users as they give a lower $R$ to larger degree users (see Figs.~\ref{fig:DegRD}d and \ref{fig:DegRD}i for GR and Figs.~\ref{fig:DegRD}e and \ref{fig:DegRD}j for IGR). In GR and IGR, the reputation is normal-like distributed and the spammers are always assigned with a low $R$. These characteristics ensure both GR and IGR owning the best performance in evaluating user reputation.

\begin{figure*}[!t]
\centering
\includegraphics[width=185mm]{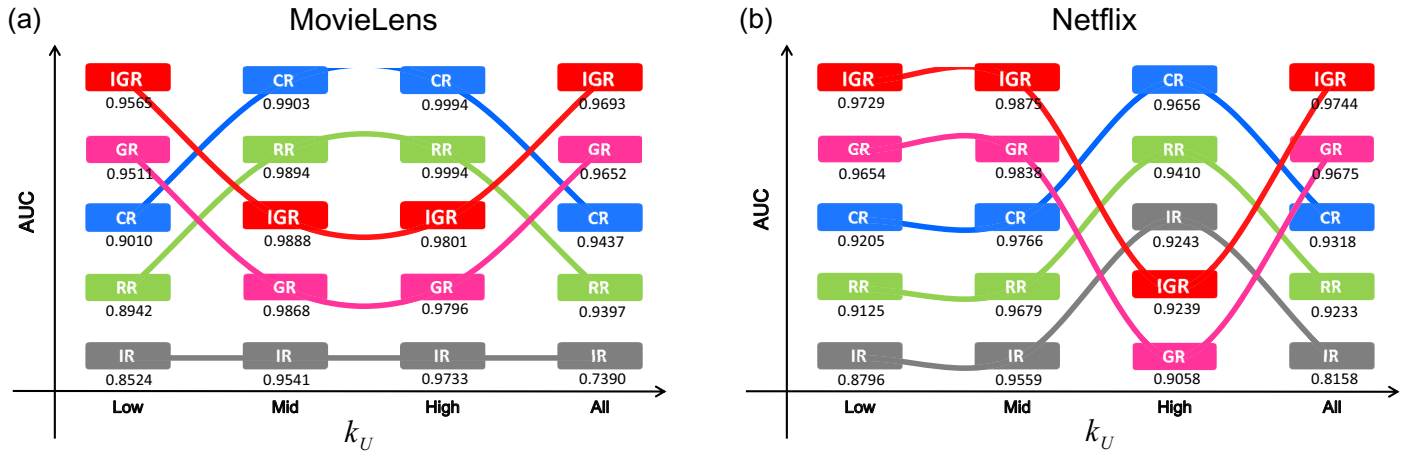}
    \caption{Comparison of difference methods in ranking random spammers with different degree $k_{U}$. Subfigures (a) and (b) are for MovieLens and Netflix, respectively. According to $k_{U}$, All users are divided into three subgroups, namely, Low, Mid and High. In each subgroup, AUC is calculated after applying different ranking methods. Accordingly, the relative ranks of these methods are obtained. The parameter is set as $p=0.1$. Results are averaged over 100 independent realizations.}
    \label{fig:RankRD}
\end{figure*}

To quantify the effects of the user degree on ranking, we divide all users into three subgroups, namely, Low, Mid and High according to their degrees. As the evidence of the heavy-tailed (i.e., stretched exponential) distribution of the user degree \cite{Shang2010}, there are only a small number of users who have large degree. To balance the number of users in each subgroups, the intervals of the user degree $k_U$ for groups Low, Mid and High are respectively set as $[k_{min}, k_{min}+0.1(k_{max}-k_{min}))$, $[k_{min}+0.1(k_{max}-k_{min}), k_{min}+0.3(k_{max}-k_{min}))$ and $[k_{min}+0.3(k_{max}-k_{min}), k_{max}]$, where $k_{min}$ and $k_{max}$ are the minimum and maximum values of $k_{U}$. In each subgroup, AUC is calculated after applying the five methods. Accordingly, the relative ranks of these methods are obtained. Results are shown in Figs.~\ref{fig:RankRD}a and \ref{fig:RankRD}b for MovieLens and Netflix, respectfully. It can be seen that IR has a limited performance for Low and Mid degree spammers. CR and GR have a good performance for High degree spammers but a poor performance for Low degree spammers. By contrast, GR and IGR outperform the other methods for Low degree spammers. In ranking All spammers, the order of these methods from the worst to the best is IR, RR, CR, GR and IGR.

\begin{figure*}[!t]
\centering
\includegraphics[width=140mm]{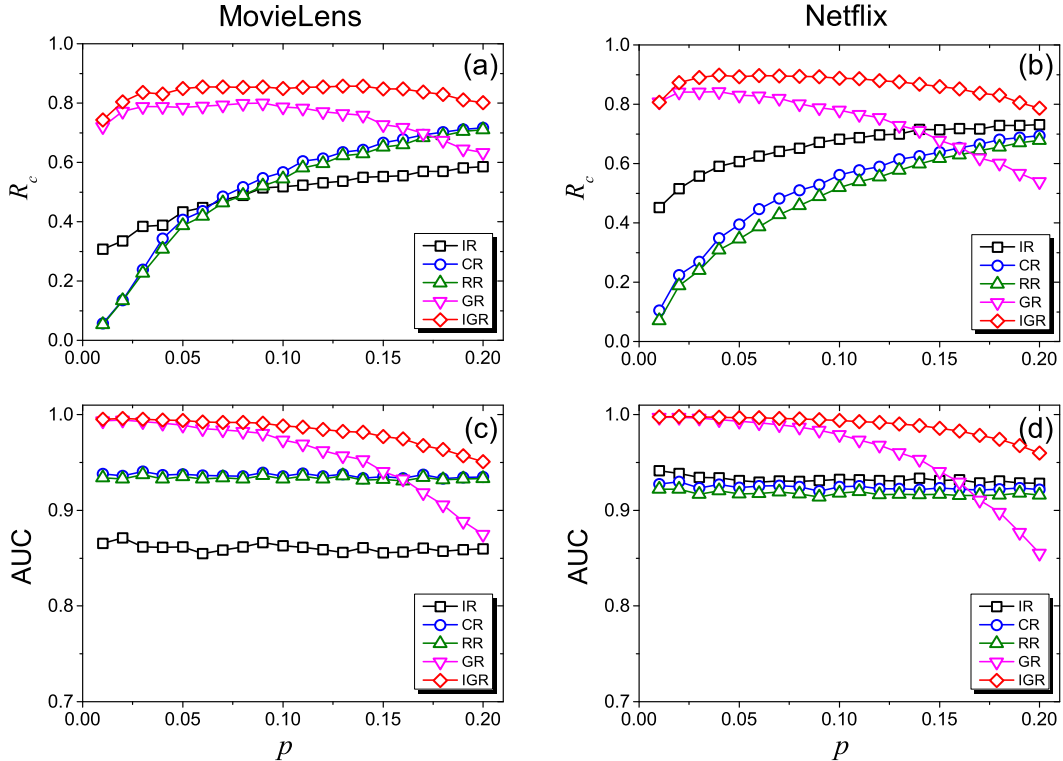}
    \caption{Performance of different methods on data sets with malicious spamming. Subfigures (a) and (b) are for $R_{c}$; subfigures (c) and (d) are for AUC. $p$ is the ratio of malicious spammers. Results are averaged over 100 independent realizations.}
    \label{fig:PerfPS}
\end{figure*}

\subsection{Malicious spamming analysis}
To evaluate the performance of different methods in resisting malicious spamming, we first generate artificial data sets with malicious spammers and then calculate $R_{c}$ and AUC accordingly. Results are shown in Fig.~\ref{fig:PerfPS}. When focusing on $R_{c}$, GR and IGR both have the best performance when the ratio of spammers $p$ is small. It is worthy noticing that IGR is much more robust than GR, since the values of $R_{c}$ in GR decrease faster than that in IGR as $p$ increases (see Figs.~\ref{fig:PerfPS}a and \ref{fig:PerfPS}b). CR and RR have the similar performance, and $R_{c}$ values in the two methods increase as $p$ increases. The performance of IR depends on the data sets, and overall it outperforms CR and RR. Moreover, we note that when $p$ is small, the values of $R_{c}$ in GR and IGR are all around 0.8, while the values in CR and RR are almost 0. These results suggest that there are some real malicious spammers in the original data sets, and GR and IGR are much better in resisting malicious spamming. Considering the overall performance indicated by AUC in Figs.~\ref{fig:PerfPS}c and \ref{fig:PerfPS}d, IGR has the best performance as the values of AUC are over 0.95. GR method is not robust than IGR especially when $p$ is large. CR and RR are robust against a large number of spammers as the AUC values are stabilized as about 0.92. Moreover, the performance of IR depends on the data sets. To conclude, in resisting malicious spamming, the group-based methods outperform the quality-based methods.

\begin{figure*}
\centering
\includegraphics[width=185mm]{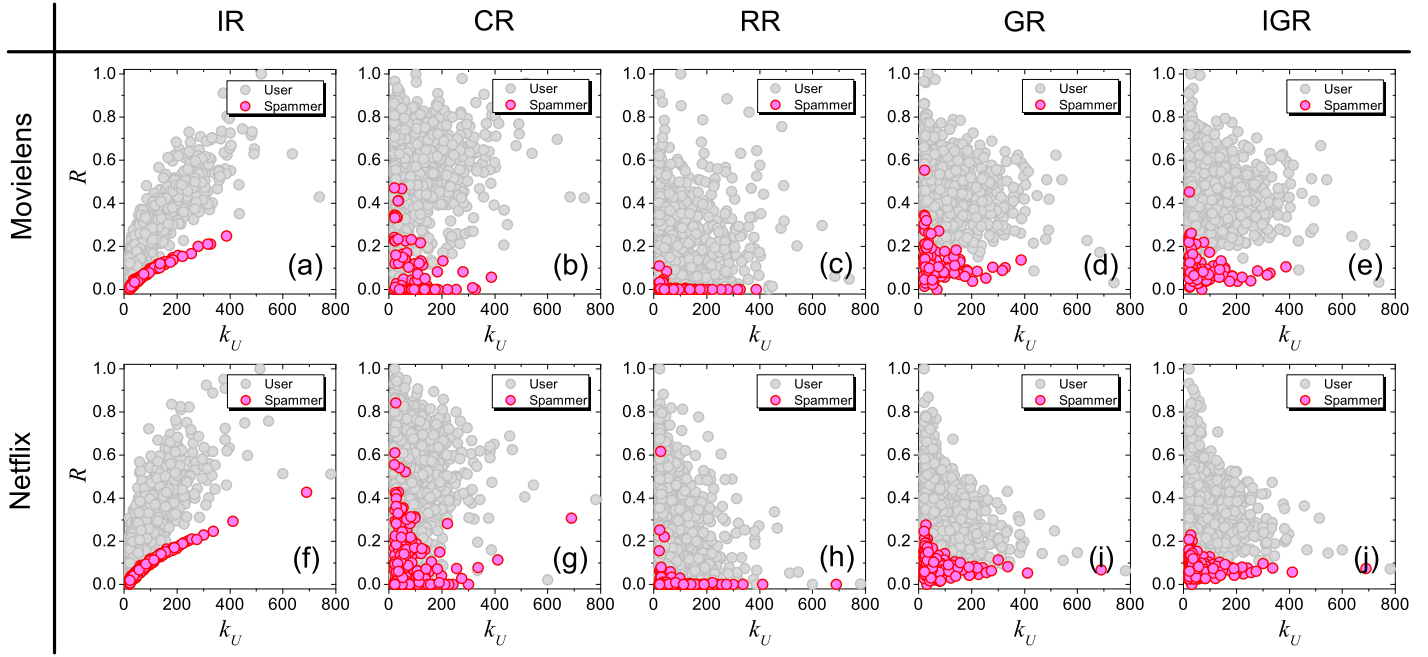}
    \caption{The relation between $R$ and $k_{U}$. $R$ is the reputation of users, obtained by applying different methods on data sets with malicious spamming. $k_{U}$ is the degree of users. The data points colored gray and pink stand for normal users and malicious spammers, respectfully. The parameter is set as $p=0.1$. Results in each subfigures are for one realization.}
    \label{fig:DegPS}
\end{figure*}

To better understand how these methods work in resisting malicious spamming, in Fig.~\ref{fig:DegPS}, we show the effect of the user degree $k_{U}$ on evaluating user reputation in parameter spaces ($R$, $k_{U}$). IR gives a high $R$ to large-degree spammers due to its preference to users of large $k_{U}$ (see Figs.~\ref{fig:DegPS}a and \ref{fig:DegPS}f). CR has no obvious degree preference as it gives high $R$ to some users regardless of their $k_{U}$ (see Figs.~\ref{fig:DegPS}b and \ref{fig:DegPS}g). RR over limits all users $R$ by giving a almost zero reputation to lots of users (see Figs.~\ref{fig:DegPS}c and \ref{fig:DegPS}h), which increases the false positive rate in spamming detection. In GR and IGR, the reputation is normal-like distributed and the spammers are always assigned with a low $R$ (see Figs.~\ref{fig:DegPS}d and \ref{fig:DegPS}i for GR and Figs.~\ref{fig:DegPS}e and \ref{fig:DegPS}j for IGR).

\begin{figure*}[!t]
\centering
\includegraphics[width=185mm]{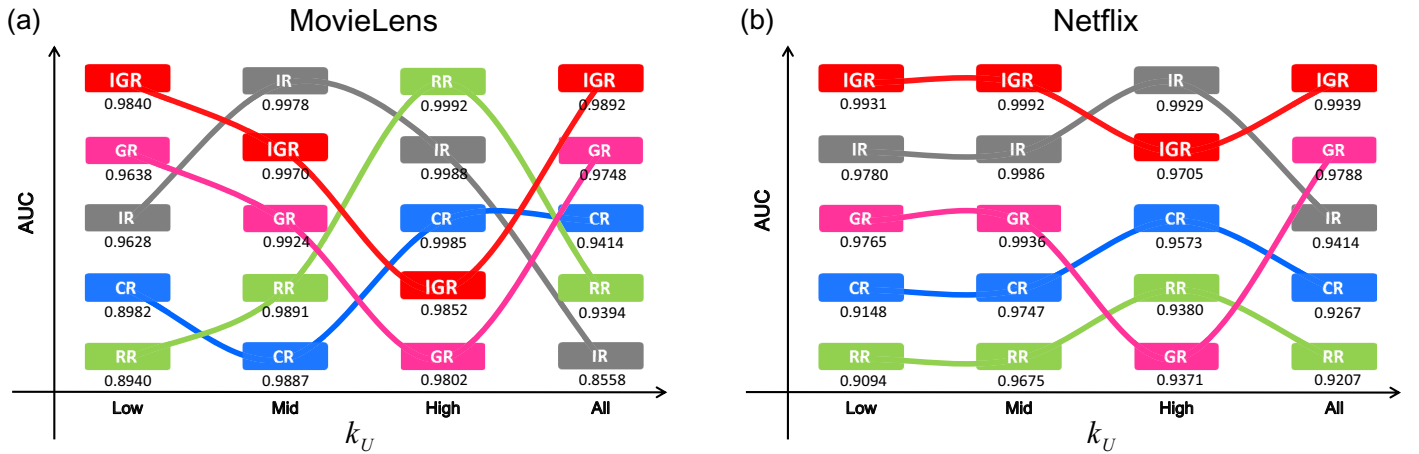}
   \caption{Comparison of difference methods in ranking malicious spammers with different degree $k_{U}$. Subfigures (a) and (b) are for MovieLens and Netflix, respectively. According to $k_{U}$, All users are divided into three subgroups, namely, Low, Mid and High. In each subgroup, AUC is calculated after applying different ranking methods. Accordingly, the relative ranks of these methods are obtained. The parameter is set as $p=0.1$. Results are averaged over 100 independent realizations.}
    \label{fig:RankPS}
\end{figure*}

To quantify the effects of the user degree $k_U$ on ranking, we show the relative ranks of different methods by AUC after dividing all users into three subgroups according to $k_{U}$ in Figs.~\ref{fig:RankPS}a and \ref{fig:RankPS}b. It can be seen that IR has better performance for Mid and High degree spammers. CR and GR perform better for High degree spammers. GR and IGR outperform the other methods for Low degree spammers although they are not competitive for High degree spammmsers. Nevertheless, in ranking All spammers, IGR again have the best performance.

\section{Conclusions and discussion}
In summary, we have proposed an iterative group-based ranking method in user reputation evaluation by introducing an iterative reputation allocation process into the original group-based ranking method. Specifically, when calculating the corresponding group sizes, ratings are assigned with higher weights if they come from users with high reputation, otherwise ratings are assigned with lower weights. In the iteration, the user reputation and the corresponding group sizes are iteratively calculated until they become stable. Extensive experiments on two real data sets suggest that the proposed method remarkably outperforms the previous quality-based methods. Further, we provided some insights on the mechanism and analyzed the characteristics of these methods. Results suggest that the iterative refinement method remarkably prefers large-degree users, the correlation-based method and reputation redistribution method have no obvious degree preference, and the group-based methods slightly prefer small-degree users.

From the macro analysis, the group-based ranking methods are distinguishable from the quality-based methods as the former ones assign users' reputation by considering their grouping behaviors while the latter ones are based on the estimation of objects' true qualities. The stability of assigning low reputation to users with high rating error and the independence of the reputation from the user degree ensure the effective of the group-based ranking methods \cite{Gao2015}. In fact, the proposed method is an improvement of the original group-based ranking method inspired by the original resource-allocation process \cite{Zhou2007bi,Ou2007} and the iterative refinement method \cite{Kleinberg1999}. In particular, compared to the original one, the proposed method is more robustness in resisting a large number of spammng attacks. That is mainly because in the proposed method the ratings from users with poor reputation have less chance in forming big groups and the reputation is iteratively updated. Even though the number of spammers increases, the effect of spam ratings on the whole system is restricted and the reputation of spammers decays through the iterations.

Our work provides a further understanding on the mechanism of some user reputation evaluation methods and gives some insights on the significance of considering users' grouping behaviors in enhancing the algorithmic performance. The proposed method is not only better in accuracy and robustness, but also easier to be implemented. Traditionally, a well-performed method should be convergent to a unique reputation vector \cite{Medo2010}, however, most of the previous reputation-based ranking methods cannot guarantee convergence \cite{Li2012ro}. Although extensive simulations suggest that the proposed method can be converge, we still expect further theoretical analysis to justify it. Moreover, the previous studies either assume a continuums of rating values such as the correlation-based method or underly the assumption of a discrete rating system such as the group-based method. In other words, how the continuous vs. discrete-valued ratings affect the user reputation evaluation is still an open issue and worth of further consideration \cite{Medo2010}. As future works, we could consider applying the proposed method to rating systems with higher-resolution scales \cite{Shi2009user} and designing more reputation evaluation methods that can make best use of users' grouping behaviors \cite{Muchnik2013}.

\section*{Acknowledgments}
This work was partially supported by the National Natural Science Foundation of China (Grant Nos. 11222543, 61370150 and 61433014). J.G. acknowledges support from Tang Lixin Education Development Foundation by UESTC. T.Z. acknowledges the Special Project of Sichuan Youth Science and Technology Innovation Research Team (Grant No. 2013TD0006).


\begin{thebibliography}{10}
\expandafter\ifx\csname url\endcsname\relax
  \def\url#1{\texttt{#1}}\fi
\expandafter\ifx\csname urlprefix\endcsname\relax\def\urlprefix{URL }\fi
\expandafter\ifx\csname href\endcsname\relax
  \def\href#1#2{#2} \def\path#1{#1}\fi

\bibitem{Resnick2000rep}
P.~Resnick, K.~Kuwabara, R.~Zeckhauser, E.~Friedman, Reputation systems,
  Commun. ACM 43~(12) (2000) 45--48.
\newblock \href {http://dx.doi.org/10.1145/355112.355122}
  {\path{doi:10.1145/355112.355122}}.

\bibitem{Standifird2001}
S.~S. Standifird, Reputation and e-commerce: ebay auctions and the asymmetrical
  impact of positive and negative ratings, J. Manag. 27~(3) (2001) 279--295.
\newblock \href {http://dx.doi.org/10.1177/014920630102700304}
  {\path{doi:10.1177/014920630102700304}}.

\bibitem{Linyuan2012}
L.~L{\"u}, M.~Medo, C.~H. Yeung, Y.-C. Zhang, Z.-K. Zhang, T.~Zhou, Recommender
  systems, Phys. Rep. 519~(1) (2012) 1 -- 49.
\newblock \href {http://dx.doi.org/10.1016/j.physrep.2012.02.006}
  {\path{doi:10.1016/j.physrep.2012.02.006}}.

\bibitem{Bobadilla2013}
J.~Bobadilla, F.~Ortega, A.~Hernando, A.~Guti{\'e}rrez, Recommender systems
  survey, Knowl.-Based Syst. 46 (2013) 109--132.
\newblock \href {http://dx.doi.org/10.1016/j.knosys.2013.03.012}
  {\path{doi:10.1016/j.knosys.2013.03.012}}.

\bibitem{Josang2007}
A.~J{\o}sang, R.~Ismail, C.~Boyd, A survey of trust and reputation systems for
  online service provision, Decis. Support Syst. 43~(2) (2007) 618 -- 644.
\newblock \href {http://dx.doi.org/10.1016/j.dss.2005.05.019}
  {\path{doi:10.1016/j.dss.2005.05.019}}.

\bibitem{Bente2012to}
G.~Bente, O.~Baptist, H.~Leuschner, To buy or not to buy: Influence of seller
  photos and reputation on buyer trust and purchase behavior, Int. J.
  Hum.-Comput. St. 70~(1) (2012) 1 -- 13.
\newblock \href {http://dx.doi.org/10.1016/j.ijhcs.2011.08.005}
  {\path{doi:10.1016/j.ijhcs.2011.08.005}}.

\bibitem{Zhao2013}
D.-D. Zhao, A.~Zeng, M.-S. Shang, J.~Gao, Long-term effects of recommendation
  on the evolution of online systems, Chin. Phys. Lett. 30~(11) (2013) 118901.
\newblock \href {http://dx.doi.org/10.1088/0256-307x/30/11/118901}
  {\path{doi:10.1088/0256-307x/30/11/118901}}.

\bibitem{Yu2015}
L.~Yu, C.~Liu, Z.-K. Zhang, Multi-linear interactive matrix factorization,
  Knowl.-Based Syst. 85 (2015) 307 -- 315.
\newblock \href {http://dx.doi.org/10.1016/j.knosys.2015.05.016}
  {\path{doi:10.1016/j.knosys.2015.05.016}}.

\bibitem{Muchnik2013}
L.~Muchnik, S.~Aral, S.~J. Taylor, Social influence bias: A randomized
  experiment, Science 341~(6146) (2013) 647--651.
\newblock \href {http://dx.doi.org/10.1126/science.1240466}
  {\path{doi:10.1126/science.1240466}}.

\bibitem{Yang2012}
Z.~Yang, Z.-K. Zhang, T.~Zhou, Anchoring bias in online voting, Europhys. Lett.
  100~(6) (2012) 68002.
\newblock \href {http://dx.doi.org/10.1209/0295-5075/100/68002}
  {\path{doi:10.1209/0295-5075/100/68002}}.

\bibitem{Toledo2015}
R.~Y. Toledo, Y.~C. Mota, L.~Mart{\'\i}nez, Correcting noisy ratings in
  collaborative recommender systems, Knowl.-Based Syst. 76 (2015) 96--108.
\newblock \href {http://dx.doi.org/10.1016/j.knosys.2014.12.011}
  {\path{doi:10.1016/j.knosys.2014.12.011}}.

\bibitem{Chirita2005}
P.-A. Chirita, W.~Nejdl, C.~Zamfir, Preventing shilling attacks in online
  recommender systems, in: Proceedings of the 7th Annual ACM International
  Workshop on Web Information and Data Management, WIDM '05, ACM, New York, NY,
  USA, 2005, pp. 67--74.
\newblock \href {http://dx.doi.org/10.1145/1097047.1097061}
  {\path{doi:10.1145/1097047.1097061}}.

\bibitem{Xie2012}
S.~Xie, G.~Wang, S.~Lin, P.~S. Yu, Review spam detection via temporal pattern
  discovery, in: Proceedings of the 18th ACM SIGKDD International Conference on
  Knowledge Discovery and Data Mining, KDD '12, ACM, New York, NY, USA, 2012,
  pp. 823--831.
\newblock \href {http://dx.doi.org/10.1145/2339530.2339662}
  {\path{doi:10.1145/2339530.2339662}}.

\bibitem{Zeng2012re}
A.~Zeng, G.~Cimini, Removing spurious interactions in complex networks, Phys.
  Rev. E 85 (2012) 036101.
\newblock \href {http://dx.doi.org/10.1103/PhysRevE.85.036101}
  {\path{doi:10.1103/PhysRevE.85.036101}}.

\bibitem{Benevenuto2009}
F.~Benevenuto, T.~Rodrigues, V.~Almeida, J.~Almeida, M.~Gon\c{c}alves,
  Detecting spammers and content promoters in online video social networks, in:
  Proceedings of the 32nd International ACM SIGIR Conference on Research and
  Development in Information Retrieval, SIGIR '09, ACM, New York, NY, USA,
  2009, pp. 620--627.
\newblock \href {http://dx.doi.org/10.1145/1571941.1572047}
  {\path{doi:10.1145/1571941.1572047}}.

\bibitem{Mukherjee2011}
A.~Mukherjee, B.~Liu, J.~Wang, N.~Glance, N.~Jindal, Detecting group review
  spam, in: Proceedings of the 20th International Conference Companion on World
  Wide Web, WWW '11, ACM, New York, NY, USA, 2011, pp. 93--94.
\newblock \href {http://dx.doi.org/10.1145/1963192.1963240}
  {\path{doi:10.1145/1963192.1963240}}.

\bibitem{Lin2014to}
Y.~Lin, T.~Zhu, X.~Wang, J.~Zhang, A.~Zhou, Towards online review spam
  detection, in: Proceedings of the Companion Publication of the 23rd
  International Conference on World Wide Web Companion, WWW Companion'14,
  International World Wide Web Conferences Steering Committee, Republic and
  Canton of Geneva, Switzerland, 2014, pp. 341--342.
\newblock \href {http://dx.doi.org/10.1145/2567948.2577293}
  {\path{doi:10.1145/2567948.2577293}}.

\bibitem{Sun2012}
Y.~Sun, Y.~Liu, Security of online reputation systems: The evolution of attacks
  and defenses, IEEE Signal Proc. Mag. 29~(2) (2012) 87--97.
\newblock \href {http://dx.doi.org/10.1109/MSP.2011.942344}
  {\path{doi:10.1109/MSP.2011.942344}}.

\bibitem{liu2014new}
H.~Liu, Z.~Hu, A.~Mian, H.~Tian, X.~Zhu, A new user similarity model to improve
  the accuracy of collaborative filtering, Knowl.-Based Syst. 56 (2014)
  156--166.
\newblock \href {http://dx.doi.org/10.1016/j.knosys.2013.11.006}
  {\path{doi:10.1016/j.knosys.2013.11.006}}.

\bibitem{Zhang2012cj}
C.-J. Zhang, A.~Zeng, Behavior patterns of online users and the effect on
  information filtering, Physica A 391~(4) (2012) 1822 -- 1830.
\newblock \href
  {http://dx.doi.org/http://dx.doi.org/10.1016/j.physa.2011.09.038}
  {\path{doi:http://dx.doi.org/10.1016/j.physa.2011.09.038}}.

\bibitem{Lim2010de}
E.-P. Lim, V.-A. Nguyen, N.~Jindal, B.~Liu, H.~W. Lauw, Detecting product
  review spammers using rating behaviors, in: Proceedings of the 19th ACM
  International Conference on Information and Knowledge Management, CIKM '10,
  ACM, New York, NY, USA, 2010, pp. 939--948.
\newblock \href {http://dx.doi.org/10.1145/1871437.1871557}
  {\path{doi:10.1145/1871437.1871557}}.

\bibitem{Ling2013a}
G.~Ling, I.~King, M.~R. Lyu, A unified framework for reputation estimation in
  online rating systems, in: Proceedings of the 23rd International Joint
  Conference on Artificial Intelligence, IJCAI '13, AAAI Press, 2013, pp.
  2670--2676.

\bibitem{Hung2012}
Y.-H. Hung, T.-L. Huang, J.-C. Hsieh, H.-J. Tsuei, C.-C. Cheng, G.-H. Tzeng,
  Online reputation management for improving marketing by using a hybrid mcdm
  model, Knowl.-Based Syst. 35 (2012) 87 -- 93.
\newblock \href {http://dx.doi.org/10.1016/j.knosys.2012.03.004}
  {\path{doi:10.1016/j.knosys.2012.03.004}}.

\bibitem{Li2012ro}
R.-H. Li, J.~X. Yu, X.~Huang, H.~Cheng, Robust reputation-based ranking on
  bipartite rating networks, in: Proceedings of the 2012 SIAM International
  Conference on Data Mining, SDM'2012, SIAM, Anaheim, California, USA, 2012,
  pp. 612--623.
\newblock \href {http://dx.doi.org/10.1137/1.9781611972825.53}
  {\path{doi:10.1137/1.9781611972825.53}}.

\bibitem{Khosravifar2012}
B.~Khosravifar, J.~Bentahar, M.~Gomrokchi, R.~Alam, Crm: An efficient trust and
  reputation model for agent computing, Knowl.-Based Syst. 30 (2012) 1 -- 16.
\newblock \href {http://dx.doi.org/10.1016/j.knosys.2011.01.004}
  {\path{doi:10.1016/j.knosys.2011.01.004}}.

\bibitem{Fujimura2003}
K.~Fujimura, T.~Nishihara, Reputation rating system based on past behavior of
  evaluators, in: Proceedings of the 4th ACM Conference on Electronic Commerce,
  EC '03, ACM, New York, NY, USA, 2003, pp. 246--247.
\newblock \href {http://dx.doi.org/10.1145/779928.779981}
  {\path{doi:10.1145/779928.779981}}.

\bibitem{Liu2015rank}
X.-L. Liu, Q.~Guo, L.~Hou, C.~Cheng, J.-G. Liu, Ranking online quality and
  reputation via the user activity, Physica A 436 (2015) 629 -- 636.
\newblock \href {http://dx.doi.org/10.1016/j.physa.2015.05.043}
  {\path{doi:10.1016/j.physa.2015.05.043}}.

\bibitem{Shang2010}
M.-S. Shang, L.~L{\"u}, Y.-C. Zhang, T.~Zhou, Empirical analysis of web-based
  user-object bipartite networks, Europhys. Lett. 90~(4) (2010) 48006.
\newblock \href {http://dx.doi.org/10.1209/0295-5075/90/48006}
  {\path{doi:10.1209/0295-5075/90/48006}}.

\bibitem{Yamamoto2004}
A.~Yamamoto, D.~Asahara, T.~Itao, S.~Tanaka, T.~Suda, Distributed pagerank: a
  distributed reputation model for open peer-to-peer network, in: Proceedings
  of the 2004 Symposium on Applications and the Internet-Workshops, 2004, pp.
  389--394.
\newblock \href {http://dx.doi.org/10.1109/SAINTW.2004.1268664}
  {\path{doi:10.1109/SAINTW.2004.1268664}}.

\bibitem{lv2011le}
L.~L{\"u}, Y.-C. Zhang, C.~H. Yeung, T.~Zhou, Leaders in social networks, the
  delicious case, PLoS ONE 6~(6) (2011) e21202.
\newblock \href {http://dx.doi.org/10.1371/journal.pone.0021202}
  {\path{doi:10.1371/journal.pone.0021202}}.

\bibitem{Zhang2007re}
Y.-C. Zhang, M.~Medo, J.~Ren, T.~Zhou, T.~Li, F.~Yang, Recommendation model
  based on opinion diffusion, Europhys. Lett. 80~(6) (2007) 68003.
\newblock \href {http://dx.doi.org/10.1209/0295-5075/80/68003}
  {\path{doi:10.1209/0295-5075/80/68003}}.

\bibitem{Zhou2007bi}
T.~Zhou, J.~Ren, M.~Medo, Y.-C. Zhang, Bipartite network projection and
  personal recommendation, Phys. Rev. E 76 (2007) 046115.
\newblock \href {http://dx.doi.org/10.1103/PhysRevE.76.046115}
  {\path{doi:10.1103/PhysRevE.76.046115}}.

\bibitem{Zhang2007he}
Y.-C. Zhang, M.~Blattner, Y.-K. Yu, Heat conduction process on community
  networks as a recommendation model, Phys. Rev. Lett. 99 (2007) 154301.
\newblock \href {http://dx.doi.org/10.1103/PhysRevLett.99.154301}
  {\path{doi:10.1103/PhysRevLett.99.154301}}.

\bibitem{Tian2012}
Y.~Tian, J.~Zhu, Learning from crowds in the presence of schools of thought,
  in: Proceedings of the 18th ACM SIGKDD International Conference on Knowledge
  Discovery and Data Mining, KDD '12, ACM, New York, NY, USA, 2012, pp.
  226--234.
\newblock \href {http://dx.doi.org/10.1145/2339530.2339571}
  {\path{doi:10.1145/2339530.2339571}}.

\bibitem{Liao2014towards}
H.~Liao, A.~Zeng, Y.-C. Zhang, Towards an objective ranking in online
  reputation systems: the effect of the rating projection, arXiv:1411.4972,
  2014.

\bibitem{Laureti2006}
P.~Laureti, L.~Moret, Y.-C. Zhang, Y.-K. Yu, Information filtering via
  iterative refinement, Europhys. Lett. 75~(6) (2006) 1006.
\newblock \href {http://dx.doi.org/10.1209/epl/i2006-10204-8}
  {\path{doi:10.1209/epl/i2006-10204-8}}.

\bibitem{De2007iterative}
C.~de~Kerchove, P.~Van~Dooren, Iterative filtering for a dynamical reputation
  system, arXiv:0711.3964, 2007.

\bibitem{Zhou2011a}
Y.-B. Zhou, T.~Lei, T.~Zhou, A robust ranking algorithm to spamming, Europhys.
  Lett. 94~(4) (2011) 48002.
\newblock \href {http://dx.doi.org/10.1209/0295-5075/94/48002}
  {\path{doi:10.1209/0295-5075/94/48002}}.

\bibitem{Liao2014}
H.~Liao, A.~Zeng, R.~Xiao, Z.-M. Ren, D.-B. Chen, Y.-C. Zhang, Ranking
  reputation and quality in online rating systems, PLoS ONE 9~(5) (2014)
  e97146.
\newblock \href {http://dx.doi.org/10.1371/journal.pone.0097146}
  {\path{doi:10.1371/journal.pone.0097146}}.

\bibitem{Allahbakhsh2015}
M.~Allahbakhsh, A.~Ignjatovic, An iterative method for calculating robust
  rating scores, IEEE T. Parall. Distr. 26~(2) (2015) 340--350.
\newblock \href {http://dx.doi.org/10.1109/TPDS.2013.215}
  {\path{doi:10.1109/TPDS.2013.215}}.

\bibitem{Gao2015}
J.~Gao, Y.-W. Dong, M.-S. Shang, S.-M. Cai, T.~Zhou, Group-based ranking method
  for online rating systems with spamming attacks, Europhys. Lett. 110~(2)
  (2015) 28003.
\newblock \href {http://dx.doi.org/10.1209/0295-5075/110/28003}
  {\path{doi:10.1209/0295-5075/110/28003}}.

\bibitem{Ou2007}
Q.~Ou, Y.-D. Jin, T.~Zhou, B.-H. Wang, B.-Q. Yin, Power-law strength-degree
  correlation from resource-allocation dynamics on weighted networks, Phys.
  Rev. E 75 (2007) 021102.
\newblock \href {http://dx.doi.org/10.1103/PhysRevE.75.021102}
  {\path{doi:10.1103/PhysRevE.75.021102}}.

\bibitem{Kleinberg1999}
J.~M. Kleinberg, Authoritative sources in a hyperlinked environment, J. ACM
  46~(5) (1999) 604--632.
\newblock \href {http://dx.doi.org/10.1145/324133.324140}
  {\path{doi:10.1145/324133.324140}}.

\bibitem{Ricci2011}
F.~Ricci, L.~Rokach, B.~Shapira, Introduction to recommender systems handbook,
  in: F.~Ricci, L.~Rokach, B.~Shapira, P.~B. Kantor (Eds.), Recommender Systems
  Handbook, Springer US, 2011, pp. 1--35.
\newblock \href {http://dx.doi.org/10.1007/978-0-387-85820-3_1}
  {\path{doi:10.1007/978-0-387-85820-3_1}}.

\bibitem{Jindal2008}
N.~Jindal, B.~Liu, Opinion spam and analysis, in: Proceedings of the 2008
  International Conference on Web Search and Data Mining, WSDM '08, ACM, New
  York, NY, USA, 2008, pp. 219--230.
\newblock \href {http://dx.doi.org/10.1145/1341531.1341560}
  {\path{doi:10.1145/1341531.1341560}}.

\bibitem{Herlocker2004}
J.~L. Herlocker, J.~A. Konstan, L.~G. Terveen, J.~T. Riedl, Evaluating
  collaborative filtering recommender systems, ACM T. Inform. Syst. 22~(1)
  (2004) 5--53.
\newblock \href {http://dx.doi.org/10.1145/963770.963772}
  {\path{doi:10.1145/963770.963772}}.

\bibitem{Hanley1982}
J.~A. Hanley, B.~J. McNeil, The meaning and use of the area under a receiver
  operating characteristic (roc) curve., Radiology 143~(1) (1982) 29--36.
\newblock \href {http://dx.doi.org/10.1148/radiology.143.1.7063747}
  {\path{doi:10.1148/radiology.143.1.7063747}}.

\bibitem{Simpson1949}
E.~H. Simpson, Measurement of diversity, Nature 163 (1949) 688.
\newblock \href {http://dx.doi.org/10.1038/163688a0}
  {\path{doi:10.1038/163688a0}}.

\bibitem{Medo2010}
M.~Medo, J.~R. Wakeling, The effect of discrete vs. continuous-valued ratings
  on reputation and ranking systems, Europhys. Lett. 91~(4) (2010) 48004.
\newblock \href {http://dx.doi.org/10.1209/0295-5075/91/48004}
  {\path{doi:10.1209/0295-5075/91/48004}}.

\bibitem{Shi2009user}
X.~Shi, J.~Zhu, R.~Cai, L.~Zhang, User grouping behavior in online forums, in:
  Proceedings of the 15th ACM SIGKDD International Conference on Knowledge
  Discovery and Data Mining, KDD '09, ACM, New York, NY, USA, 2009, pp.
  777--786.
\newblock \href {http://dx.doi.org/10.1145/1557019.1557105}
  {\path{doi:10.1145/1557019.1557105}}.

\end{thebibliography}
\end{document}